\documentclass[a4paper,fleqn,usenatbib]{mnras}

\usepackage{mathptmx}

\usepackage[T1]{fontenc}
\usepackage{ae,aecompl}

\usepackage{graphicx}	
\usepackage{amsmath}	
\usepackage{amssymb}	
\usepackage{rotating}

\title{On the robustness of the H$\beta$ Lick index as a cosmic clock in passive early-type galaxies}

\author[A. Concas et al.]{
Alice Concas,$^{1,2,3}$\thanks{E-mail: alice.concas@tum.de (KTS)}
L. Pozzetti,$^{3}$
M. Moresco$^{2,3}$
and A. Cimatti$^{2}$
\\
$^{1}$Excellence Cluster Universe, Boltzmannstr. 2  D-85748 Garching,  Germany \\
$^{2}$Dipartimento di Fisica e Astronomia, Universit\'a degli Studi di Bologna, V.le Berti Pichat, 6/2 - 40127, Bologna, Italy\\
$^{3}$INAF - Osservatorio Astronomico di Bologna, via Ranzani 1, I-40127, Bologna, Italy
}

\date{Accepted XXX. Received YYY; in original form ZZZ}

\pubyear{2016}

\begin{document}

\label{firstpage}
\pagerange{\pageref{firstpage}--\pageref{lastpage}}
\maketitle

\begin{abstract}

{We examine the H$\beta$ Lick index in a sample of $\sim 24000$ massive ($\rm log(M/M_{\odot})>10.75$) and passive early-type galaxies extracted from SDSS at z<0.3, in order to assess the reliability of this index to constrain the epoch of formation and age evolution of these systems. We further investigate the possibility of exploiting this index as "cosmic chronometer", i.e. to derive the Hubble parameter from its differential evolution with redshift, hence constraining cosmological models independently of other probes. We find that the H$\beta$ strength increases with redshift as expected in passive evolution models, and shows at each redshift weaker values in more massive galaxies. However, a detailed comparison of the observed index with the predictions of stellar population synthesis models highlights a significant tension, with the observed index being systematically lower than expected. By analyzing the stacked spectra, we find a weak [NII]$\lambda6584$ emission line (not detectable in the single spectra) which anti-correlates with the mass, that can be interpreted as a hint of the presence of ionized gas. We estimated the correction of the H$\beta$ index by the residual emission component exploiting different approaches, but find it very uncertain and model-dependent. We conclude that, while the qualitative trends of the observed H$\beta$-z relations are consistent with the expected passive and downsizing scenario, the possible presence of ionized gas even in the most massive and passive galaxies prevents to use this index for a quantitative estimate of the age evolution and for cosmological applications.}
\end{abstract}





\section{Introduction}\label{introduzione}
Early-type galaxies (ETGs) are perfect candidates to investigate both galaxy formation and evolution theories and, at the same time, to put constraints on the expansion history of the Universe. Many recent observations show that most massive galaxies contain the oldest stellar populations up to 
$z\sim 1-2$ \citep[e.g.][]{Spinrad1997,Cowie1999,Heavens2004,Thomas2005,Cimatti2008,Thomas2010,Moresco2011} and that only few percent of their current stellar mass was formed after $z \sim 1$.{ Moreover, studies on the stellar mass function at different redshifts \citep[e.g.][]{Bundy2005,Borch2006,Bundy2006,Ilbert2010,Pozzetti2010,Maraston2013} showed that most massive galaxies ( $M \sim 10^{11} M_{\odot}$) are characterized by an increase of their stellar mass density at z$\sim$1(e.g., \citealp{Ilbert2010} based on the COSMOS 2 deg$^2$ field) and a subsequent negligible evolution in number density from $z \sim 1$ to the present \citep{Cimatti2006,Pozzetti2010,Moresco2013}.
All these observational evidences suggest that these massive, red and passively evolving galaxies represent the oldest objects in the Universe below z$\sim 1$.}
For this reason, a study based on the variation of ETGs properties, such as age, chemical composition, dynamic, at different redshifts, could rightly 
help to expand our current knowledge of the physical process that drive the formation and evolution of galaxies.
Moreover, as firstly pointed out by \cite{Jimenez2002}, the differential age evolution of ETGs can be also used to put several cosmological constrains. 
In particular, by estimating the relative differential age ($\Delta t_{age}$) between ETGs, formed at the same cosmic time,  but observed at different redshifts ($\Delta z$), it is possible to determine the derivative of redshift with respect to cosmic time $dz/dt$.

This quantity is directly related to the cosmological expansion history described by the Hubble parameter: $ H(z)=- \frac{1}{1+z} \frac{dz}{dt}$, which can be used to estimate many cosmological parameters such as $H_0$, $\Omega_{0,M}$, $\Omega_{0,\Lambda}$ and $w_{\Lambda}$. This technique, known as ``cosmic chronometers'' method, has been used recently in several works and obtained promising results in the estimate of the Hubble parameter and of the dark energy equation of state \citep{Simon2005,Stern2010,Moresco2012a,Moresco2012b,Zhang2012,Moresco2016b}. The basic problem of this method is the fact that it relies on the measure of stellar population age, which presents, when estimated from a fit of the spectrum and/or spectral energy distribution (SED), significant degenerations with other parameters, such as metallicity, star formation history (SFH) and dust content. An alternative method to mitigate these degeneracies is to identify a particular spectral feature that is sensitive to the aging of the stellar population. An example of this approach is provided by \cite{Moresco2011,Moresco2012a}, \cite{Moresco2015} and \cite{Moresco2016a}, in which the expansion history of the Universe is determined by studying the redshift dependence of the $4000$ \AA\ break. Despite the high potential of the method, the break at $4000$ \AA\ is particularly sensitive not only to the age, but also to the variation of the metallicity of the stellar population, which has to be taken into account as a systematic error in the estimate of cosmological parameters.
It is evident the need to find a new spectral feature that minimize the dependence systematic and able to translate the amount $dz/dt$ in the observable $dz/d \textrm{(feature)}$ and then able to determine the cosmological parameter with higher accuracy.

In the effort of exploiting reliable age indicators, we present a new study of ETGs relative age evolution based on the variation of the H$\beta$ Lick index with redshift. This index, firstly introduced by \cite{Burstein1984} and redefined by \cite{Worthey1994}, has been identified as the one with the maximum dependence on the variation of the age of the stellar population, and the minimal dependence 
on metallicity and chemical composition \citep{Trager1998,Lee2005,Thomas2011}. 
In detail, we measure the H$\beta$ Lick index in a sample of passive galaxies selected from the Sloan Digital Sky Survey (SDSS), to check if this particular spectral feature is actually able to provide a reliable differential dating of galaxy stellar ages, in order to derive constraints both on galaxy evolution and formation and on the expansion history of the Universe. \\
This paper is organized as follows. In Section \ref{sample} we introduce the data sample. We describe the physical parameters of the passive galaxies sample, and our method for the H$\beta$ Lick measurement. In Section \ref{results} we present the results about the H$\beta -$z  and H$\beta-$mass relations and compare them with the theoretical predictions of the stellar population synthesis models. We discuss the effect of possible contamination by the emission lines in the observed H$\beta$ absorption line in the Section \ref{discussion}, and finally in Section \ref{conclusioni} we list our conclusions. Throughout the paper we assume a $H_0=70$ $km\,s^{-1}Mpc^{-1}$, $\Omega_{0,M}=0.25$ and $\Omega_{\Lambda}=0.75$ cosmology.


\section{Data}  \label{sample}

The galaxies analyzed in this work are selected from the spectroscopic catalogue of the Sixth Data Release of Sloan Digital Sky Survey (SDSS-DR6). 
The SDSS-DR6 provides photometry in the u, g, r, i, z bands and the spectrum of about $800000$ galaxies, extracted with Petrosian magnitude $r<17.77$. 
Each spectrum is measured from $3800$ to $9200$ \AA\ with a resolution $\lambda /\Delta \lambda  \approx 1800-2000$\footnote{http://www.sdss3.org/dr9/scope.php\# specstats}.
The SDSS data are provided in vacuum wavelengths. To be meaningfully compared with atomic transitions, we converted them in the air system\footnote{The conversion relation from vacuum wavelengths to air wavelengths, proposed in \cite{Morton1991}, is the following:
\begin{equation}
\lambda_{AIR}= \dfrac{\lambda_{VAC}}{\left(  1.0 + 2.735182E-4 + \dfrac{131.4182}{ \lambda_{VAC}^2}  + \dfrac{2.76249E8}{ \lambda_{VAC}^4}  \right)} 
\end{equation} with $\lambda_{AIR}$ and $\lambda_{VAC}$ expressed in \AA.\\}.
To have a wider photometric coverage, we decided to analyze the sample provided by \cite{Moresco2011}, where SDSS galaxies are matched with the Two Micron All Sky Survey (2MASS), with photometry in the J, H and K bands.
This wider coverage allows a better accuracy in the stellar mass estimates from the SED-fitting, and a more precise selection based on photometric data (see Sec. \ref{selection}).

\subsection{The sample of passive ETGs}
\label{selection}

This study strongly relies on the selection of a sample of passively evolving galaxies which, having assembled most of their mass at high redshifts, are able to trace homogeneously the age-evolution of the Universe. To select the purest possible sample of passive galaxies, with no evidence of recent episodes of star formation, we followed the approach described in \cite{Moresco2011}, and selected galaxies with:
\begin{itemize}
\item {\bf no strong emission lines.} Starting from the equivalent width measurements provided by the analysis of the MPA-JHU 
DR7\footnote{http://www.mpa-garching.mpg.de/SDSS/DR7/}, we select only the objects with a restframe equivalent width EW(H$\alpha$)>-5 \AA\ and EW([OII]$\lambda$3727)$>$
-5~\AA, where, by convention, the measures are given in negative values when in emission.
\item {\bf spectral energy distribution matching the reddest passive ETG templates.} Following the approach of \cite{Zucca2006}, the SED of each galaxy has been compared with a library of 62 empiric SEDs described in \cite{Ilbert2006}, comprising galaxies with both old stellar populations and intense star formation activity.
The galaxies are divided according to their SED, and we select those galaxies matching an early-type template, corresponding to the four reddest templates.
\end{itemize}
With this selection, we obtain a sample of $\sim 100000$ galaxies. For more details, we refer to \cite{Moresco2011}.

\begin{figure}
   \centering
   \includegraphics[angle=270,width=1.\hsize]{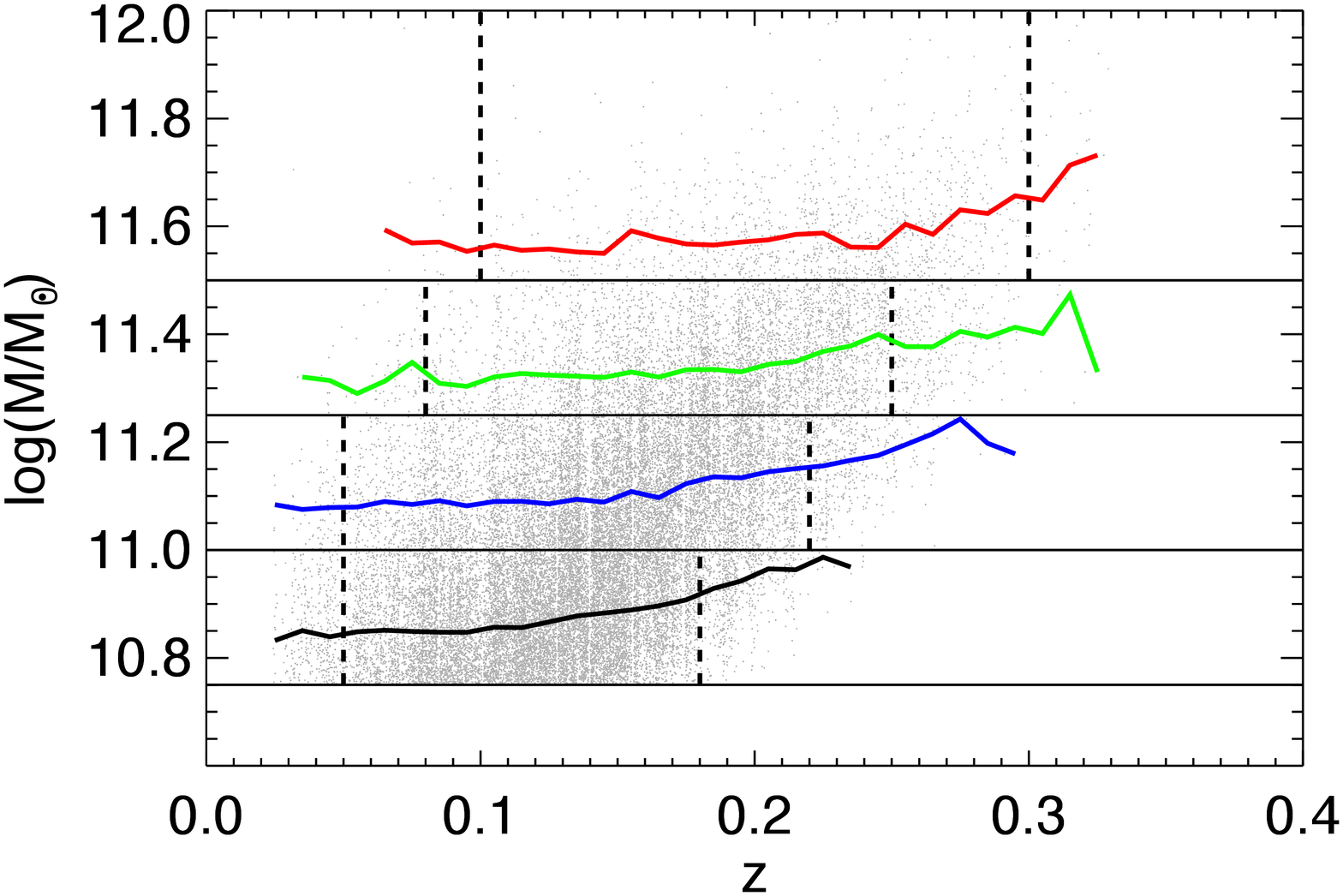}
   \caption{Stellar mass$-$redshift distributions of the four mass subsamples (red, green, blue and black curves respectively for the BinIV, BinIII, BinII and BinI, from top to bottom). The black dotted vertical lines indicate the redshift division of the sample (see Tab. \ref{tab:InfoBin} for the stellar mass median values). {(A color version of this figure is available in the online journal.)}}
              \label{FigMassa}%
              \end{figure}

In order to investigate the evolution of our sample, we study their physical properties, focusing in particular on the stellar mass, stellar metallicity and velocity dispersion.
We use the stellar mass estimates from \cite{Moresco2011}, obtained by a multi-colour SED-fitting procedure ($8$ photometric bands from u to K).
We apply a further mass cut, selecting only the most massive galaxies with a stellar mass $\log \left ( M/M_{\odot} \right) \geq 10.75 $.
This choice is made to select only the most massive objects that, according to the downsizing scenario, should correspond to the oldest objects \citep{Thomas2010}. 
In the {\it mass-downsizing} scenario, the most massive galaxies were formed earlier and over a shorter period than those in less massive galaxies. For this reason, we decide to split our sample in four narrow mass bins ($\Delta \log(M/M_{\odot}) =0.25$), to have a more homogeneous sampling of the redshift of formation:
$10.75 < \log(M/M_{\odot}) < 11 $ (BinI), $11 < \log(M/M_{\odot}) < 11.25 $ (BinII), $ 11.25 < \log(M/M_{\odot}) < 11.5 $ (BinIII) and $\log(M/M_{\odot}) > 11.5 $ 
(BinIV) (see Fig. \ref{FigMassa}).
Furthermore, for this sample we have a measure of the Star Formation History (SFH) estimated by \cite{Moresco2011} with an exponential delayed SFH: $SFR(t,\tau) \propto \left ( t/\tau^{2} \right) \exp \left(-t/\tau \right)$ (with $\tau$ in the range of values $0.05 \leq \tau \leq 1$ Gyrs). The median value of the our total sample is $\tau = 0.2 $ Gyrs. 



In Fig. \ref{FigMassa} we present the mass-redshift distribution.
It is evident that the median mass increases with the redshift due to magnitude limit effect of the survey.
Therefore, to avoid possible biases, we select the galaxies having $z\leq 0.18$, $z \leq 0.22$, $z \leq 0.25$ and $z \leq 0.3$, respectively for the BinI, BinII, BinIII and BinIV. 
In this way, we limit the difference in the median mass along the redshift range to $0.1$ dex on average (see Fig. \ref{FigMassa}).

As mentioned in the previous section, the H$\beta$ Lick index is the less sensitive to metallicity variations compared to all other indices of the Lick system. However, in the wavelength range where the index is defined, there may be weak metal absorption lines and, for this reason, the metallicity dependence of the $H\beta$ index is not totally negligible. 
The sample has been therefore cross-matched with the SDSS-DR4 subsample obtained by \cite{Gallazzi2005}, for which the stellar metallicities of the galaxies are estimated from the simultaneous fit of several spectral features (D4000, H$\beta$, H$\delta_{a}+$H$\gamma_{a}$, [Mg$_{2}$Fe] and [MgFe]'). 
We study the metallicity-redshift relation. In order to minimize the possible metallicity effect, 
we made a further cut in redshift, selecting only galaxies with redshift $z\geq 0.05$ for BinI and BinII, $z \geq 0.08$ for BinIII, and $z \geq 0.1$ for BinIV.
{The median metallicity of the final sample is $Z/Z_{\odot}\sim1.2$ (with a total spread of about $0.3$ within the total redshift range analyzed and of $\sim$0.2 in narrow redshift bins $\Delta z \sim 0.02-0.03$), showing only a slight variation with redshift, in agreement with the recent metallicity measurements obtained by total spectral fitting of our stacked spectra (\citealp{Citro2015}). 
We note, further, that there is no significant difference in the median metallicity of the different mass subsamples.}

%
%
%



\begin{figure}
       \includegraphics[angle=-90,width=\hsize]{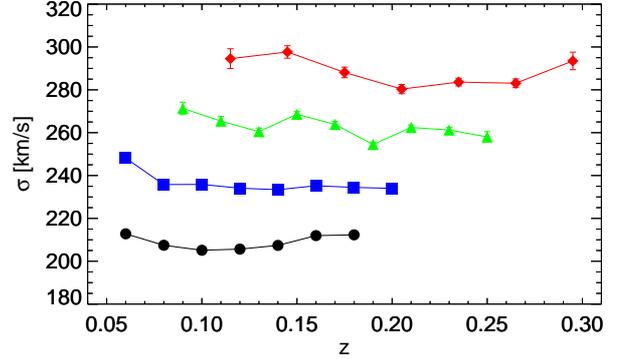}
   \caption{Velocity dispersion-redshift relations, averaged in bin of redshift, for the different mass subsamples, BinIV, BinIII, BinII and BinI from top to bottom (see Tab. \ref{tab:InfoBin} for the median values). {(A color version of this figure is available in the online journal.)}} 
             \label{FigSigma}%
    \end{figure}

In this paper, we adopt the velocity dispersion derived by the Princeton group\footnote{
The data are available on the website http://spectro.princeton.edu/}.
Figure \ref{FigSigma} shows the velocity dispersion relations for all mass subsamples.
We find that the velocity dispersion is constant in each subsample (see Tab \ref{tab:InfoBin} for the median values).
    
Finally, the observed optical spectrum can be contaminated by the presence of night sky emission lines \citep[see][]{Hanuschik2003}.
Particularly the night sky emission lines at $5577$ \AA\ ([OI]) and at $5889.950$, $5895.920$ (NaI) \AA\  are very strong, and leave significant contamination in H$\beta$ measurement.

In order to avoid this source of contamination, we decide to excluded the galaxies having a redshift between $0.14<z<0.155$ and $0.20<z<0.22$, which are the redshifts for which those lines fall into the H$\beta$ definition range.

\begin{table*}
\small
\centering
\begin{tabular}{l|c|c|c|c|c}

\hline
\hline
     & redshift        & mass          & median        &  median         &  $\#$ galaxies \\
     &    z            & range         & mass           &  velocity dispersion         &                 \\
     &                 & $\log(M/M_{\odot})$  & $\log(M/M_{\odot})$     & $\sigma$ [km s$^{-1}]$   &   \\
\hline
BinI  &$0.05-0.18$     & $10.75-11.0$         & $10.87 \pm 0.001$       & $207.3 \pm 0.3$& $11631$        \\
BinII &$0.05-0.22$     & $11.0-11.25$        & $11.10 \pm 0.001$     & $234.5 \pm 0.4$& $7990$         \\
BinIII&$0.08-0.25$     & $11.25-11.5$       & $11.34 \pm 0.001$      & $259.9 \pm 0.6$& $3314$          \\
BinIV &$0.1-0.3$       & $>11.5$             & $11.58 \pm 0.002$       & $286.1 \pm 1.1$& $979$          \\
\hline
\hline
Tot  &$0.05-0.3$       & $>10.75$             & $11.01 \pm 0.001$      & $226.7 \pm 0.2$ & $23914$          \\
\hline
\hline
\end{tabular}
\caption{Redshift range, mass range, median mass and median velocity dispersion of the passive galaxies in different mass subsamples.}
\label{tab:InfoBin}
\end{table*}
\normalsize

       \begin{figure*}
\includegraphics[angle=-90, width=\hsize, trim = 55mm 10mm 0mm -1mm, clip=true]{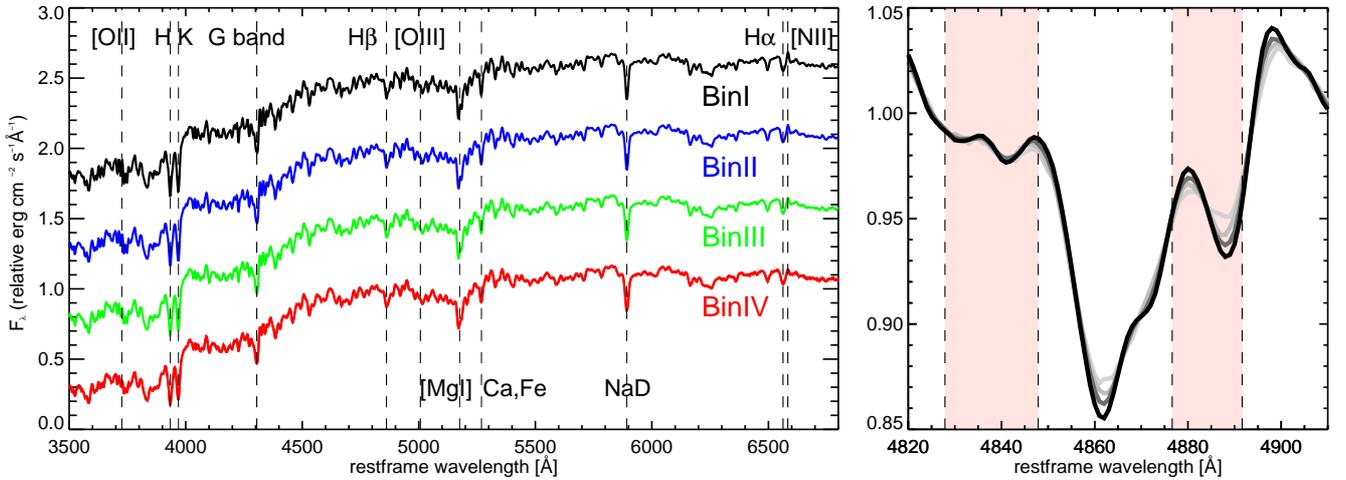}
   \caption{{Median stacked spectra of the four passive galaxies mass subsamples. (from top to bottom respectively for the BinI, BinII, BinIII and BinIV, left panel). In the range between $3500-6800$ \AA\, there are not significant emission lines ([OII], [OIII], H$\alpha$ and [NII]$\lambda$6584 ). The H$\beta$ line show a typical absorption feature (right panel) in all the mass bins (from black to light gray for BinI, BinII, BinIII and BinIV respectively). The spectra are normalized between $5000$ and $5500$ \AA\ restframe. }{(A color version of this figure is available in the online journal.)}}
              \label{spettri}%
    \end{figure*}
 
After all the cuts adopted, the final sample consists of $\sim 23914$ galaxies. 
The median values and the respective errors \footnote{The error on the median are evaluated as the $median$ $absolute$ $deviation$ $/\sqrt N$, $MAD = 1,482 * median 
(| x - median (x) |)$ \citep[see][]{Hoaglin1983}.}; of mass and velocity dispersion for the different mass subsamples are reported in Table \ref{tab:InfoBin}.
In the same table, we also report the number of galaxies in each bin.

Fig. \ref{spettri} shows the median stacked spectra for the four mass subsamples.
From the left panel, it is evident that all spectra show typical features of a passive population, with (i) absence of emission lines ( [OII], H$\beta$, [OIII], and H$\alpha$), (ii) strong absorption features (H$\beta$, H$\alpha$, G band, Ca H and K), and (iii) strong D4000 break.
We notice that there is still present a weak emission at $6584$ \AA\ [NII], and that the amplitude of this feature is mass-dependent, being most evident in the less massive subsample (BinI).
For more details, see Section \ref{discussion}.
{We highlight the region near H$\beta$ line (right panel); all the four spectra show a typical and unperturbed absorption profile.}
    
\subsection{H$\beta$ index}\label{index}

We measure the H$\beta$ Lick absorption line index for all $23914$ galaxies.
We implement an IDL (Interactive Data Language\footnote{http://www.exelisvis.com/ProductsServices/IDL.aspx}) code in order to perform measures coherently on observational data and models.
We use the Lick H$\beta$ index definition by \cite{Worthey1994}; the bandpasses are shown in Fig. \ref{spettri}.
In the standard Lick system, the indices are defined on low resolution spectra ($\sim9$ \AA\ FWHM in H$\beta$ region). However, the modern spectroscopic surveys, and in particular the SDSS, provide spectra with a much higher resolution ($\sim 2.4-2.7$ \AA\ FWHM), comparable with the new stellar populations synthesis models (for example \citealp{Bruzual2003} and \citealp{Maraston2011} ).
In order to measure coherently this feature in the models and observations, we decide to perform the analysis directly on the observed spectra, without altering the resolution.

Before applying our code to real data, we test its reliability to reproduce the H$\beta$ Lick values tabulated in the BC03 templates. We find a very good agreement between the two measurements with a negligible median offset of $2$ per cent. 
We also compare our measurements with those derived by the MPA-JHU group for the same galaxy sample (http://www.mpa-garching.mpg.de/SDSS/DR7/).
We find a good agreement between the values, with a mean difference of $-0.03$ \AA\, and a standard deviation of $0.033$ \AA\ , which is comparable with the average measured error.
Therefore, we find that the two independent methods are consistent, and without any significant biases.


\section{Trends of H$\beta$} \label{results}
We derive the median H$\beta$-redshift relations in each mass subsample; 
we use narrow redshift bins ($\Delta z \sim 0.02-0.03$) for the three most populated mass subsamples, while in the more massive subsample we use $\Delta z \sim 0.04$ because of the lower statistic with respect to the other samples.

\begin{figure}
\centering
\includegraphics[angle=-90,width=\hsize]{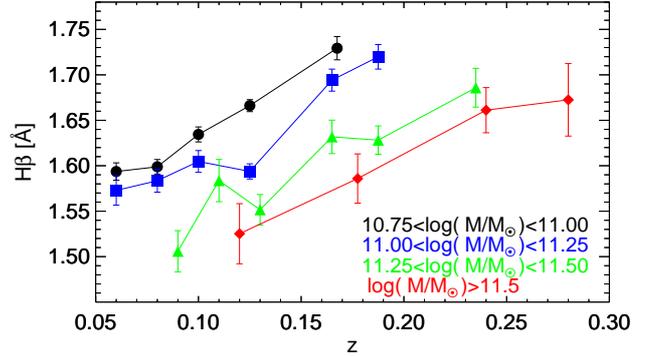}
\caption{{Median H$\beta$-redshift relation for passive galaxies in different mass subsamples (circles, squares, triangles and diamonds for BinI, BinII, BinIII and BinIV respectively). (A color version of this figure is available in the online journal.) }}
\label{fig:HbetaMediano}
\end{figure}

The results are shown in Fig. \ref{fig:HbetaMediano}. 
From this figure, we note two main trends:
\begin{itemize}
\item {a clear H$\beta-$redshift relation for each subsample, with an increase in the index strength of $\sim$10\% along the redshift range.}
\item{a clear trend with mass; i.e., at each redshift more massive galaxies present a lower H$\beta$ index with respect to less massive ones ($\sim 9 \%$).}
\end{itemize}

\begin{figure*}
\centering
\includegraphics[angle=-90, width=\hsize]{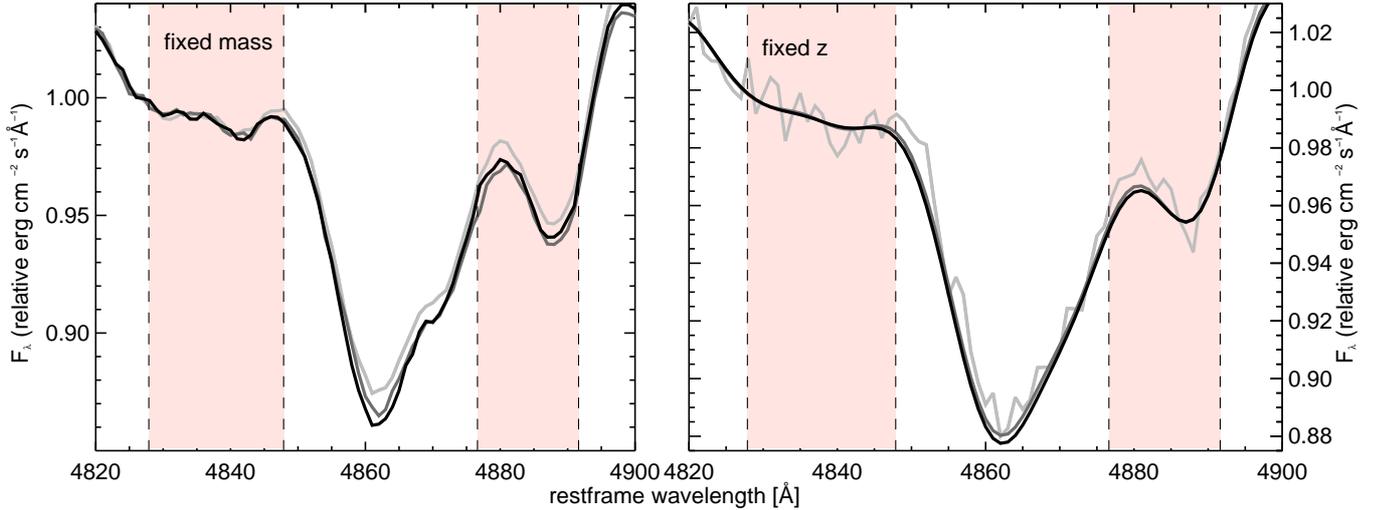}
\caption{Median stacked spectra in the H$\beta$ line region. {Left panel: stacked spectra with the same mass ($11 < \log(M/M_{\odot}) < 11.25$, BinII) and different mean redshift $z=0.08$, $0.10$ and $0.1875$} (from light gray to black, respectively). Right panel: stacked spectra with the same redshift, $z=0.125$, for different mass subsamples { (black, gray and light gray for BinI, BinII, and BinIV respectively). For illustrative purposes the BinI and BinII spectra are smoothed to the common velocity dispersion of BinIV.  }}
\label{fig:SED}
\end{figure*}

It is interesting to note that both effects can easily be seen even by direct inspection of the median spectra.
As an example, we show the stacked spectra at different redshifts for the sample with $11 < \log(M/M_{\odot}) < 11.25 $ (BinII, see Fig. \ref{fig:SED}, left panel), from which it is evident a deeper absorption line with redshift. 
The trend with the mass is also visible in the stacked spectra of galaxies at fixed redshift, with most massive galaxies showing a smaller absorption line than lower-mass ones (see Fig. \ref{fig:SED}, right panel).

Both these trends are qualitatively consistent with a general aging of the stellar population and a mass-age relation (``mass-downsizing''), also found by \cite{Thomas2010}.
Our results also concur with a recent work of \cite{Choi2014}. By modeling the full optical spectrum of a quiescent galaxy sample at low and intermediate redshift ($0.07<z<0.09$ from the SDSS and $0.1<z<0.7$ from AGN and Galaxy Evolution Survey, AGES), \cite{Choi2014} found that the best-fit SSP-equivalent age is higher for the most massive galaxies and it increases with decreasing redshift at fixed stellar mass.
 This correspondence between two different methods highlights the potential of our H$\beta$ line as a tracer of aging in the stellar population.
We test this interpretation in Sec. \ref{check-results}.

\subsection{Sensitivity to metallicity and velocity dispersion}
\label{check-results}

The H$\beta$ index is not totally immune to metallicity ($Z/Z_{\odot}$) effects for the integrated stellar population (\citealp{Worthey1994}). Furthermore, this index is sensitive to the galaxy velocity dispersion ($\sigma$).
To exclude the possibility that the trends found are due to a $\sigma$ or $Z/Z_{\odot}$ variation, we study the impact of these parameters using stellar population synthesis  (SPS) models.
We consider \cite{Bruzual2003} and \cite{Maraston2011} models (hereafter BC03 and MaStro, respectively).
BC03 models are available at a resolution of $3$ \AA\ FWHM in the wavelength range between $3200-9500$~\AA; MaStro models are available with a resolution of $\sim2.3$ \AA\ FWHM  \footnote{More precisely, in \cite{Beifiori2011} it is shown that the spectral resolution of the MaStro models, based on the library stellar MILES is about  $2.54$ \AA\ FWHM, instead of the nominal $2.3$ \AA\ } between $3525-7500$~\AA\ . Both models share a resolution very similar to the one of SDSS spectra ($\approx1800-2000$ between $3800$ to $9200$~\AA). This is particularly useful since it allows a direct comparison between the observed spectra with the theoretical SED without having to modify the spectral resolution.

These models are used to create libraries of synthetic spectra with different velocity dispersions ($\sigma=0, 200, 250$ and $300$ km s$^{-1}$) and different metallicity ($Z/Z_{\odot}=1,1.4,1.5$ obtained by interpolating quadratically between $Z/Z_{\odot}=0.4, 1, 2.5$ available in BC03 and $Z/Z_{\odot}=0.5, 1, 2$ in MaStro).
The grid has been created with age $0.5 \leq t \leq 15 $ Gyr, with delayed exponential SFH, SFR$(t,\tau) \propto (t/{\tau}^{2})\exp(-t/\tau)$, where $\tau$ is chosen in the range $0.05 \leq \tau \leq 1$ Gyr.

\subsubsection{Effects of the velocity dispersion}
To check the effect of the  velocity dispersion in our H$\beta$ Lick measures, we perform several tests.
By using the theoretical spectra described in the previous section, we verify that an increase of velocity dispersion causes an apparent decrease of the observed H$\beta$ index strength only due to a broadening effect.
We further estimate the maximum percentage difference in H$\beta$ between our mass bins 
which can be attributed to $\sigma$ from the analysis of SPS models; we find that 
the percentage difference in H$\beta$ between a single stellar population (SSP) with $\sigma=200$ km s$^{-1}$ and $\sigma=250$ km s$^{-1}$ is $<1.5\%$, and is $< 3\%$ between $\sigma=200$ km s$^{-1}$and $\sigma=300$ km s$^{-1}$, for both models.
We underline that given the range of $\sigma$ of our data, these two estimates represent the maximum  difference in H$\beta$ which can be attributed to a $\sigma$ effect between our mass subsamples, and they are always lower than the mean percentage difference in our subsamples with similar $\sigma$ differences.
Indeed, the mean percentage difference in H$\beta$ median values 
between BinI and BinIII is $\sim$6\%, and $\sim$9\% between BinI and BinIV (similarly, we get a difference of $\sim$ 4\% between BinII and BinIII, and $\sim$6\% between BinII and BinIV). 
Moreover, we note that all SPS models analyzed show that SEDs with different velocity dispersions have a difference in flux in the red pseudo-continuum region ($4876.625-4891.625$ ), but the stacked spectra of the same mass subsample do not show this trend (see Fig. \ref{fig:SED}).

For this reason, we exclude that the velocity dispersion by itself can cause the observed mass segregation between the H$\beta$-z relation in different mass subsamples.

Finally, we verified that in our data the median velocity dispersion (see Table \ref{tab:InfoBin} for median values
of BinI, BinII, BinIII and BinIV) shows
almost no redshift evolution in each mass bin (see Fig. \ref{FigSigma}). For this reason, we exclude that the observed H$\beta$ index increase with redshift can be due to a velocity dispersion effect.



\begin{figure*}
\centering
\includegraphics[angle=-90, width=\hsize]{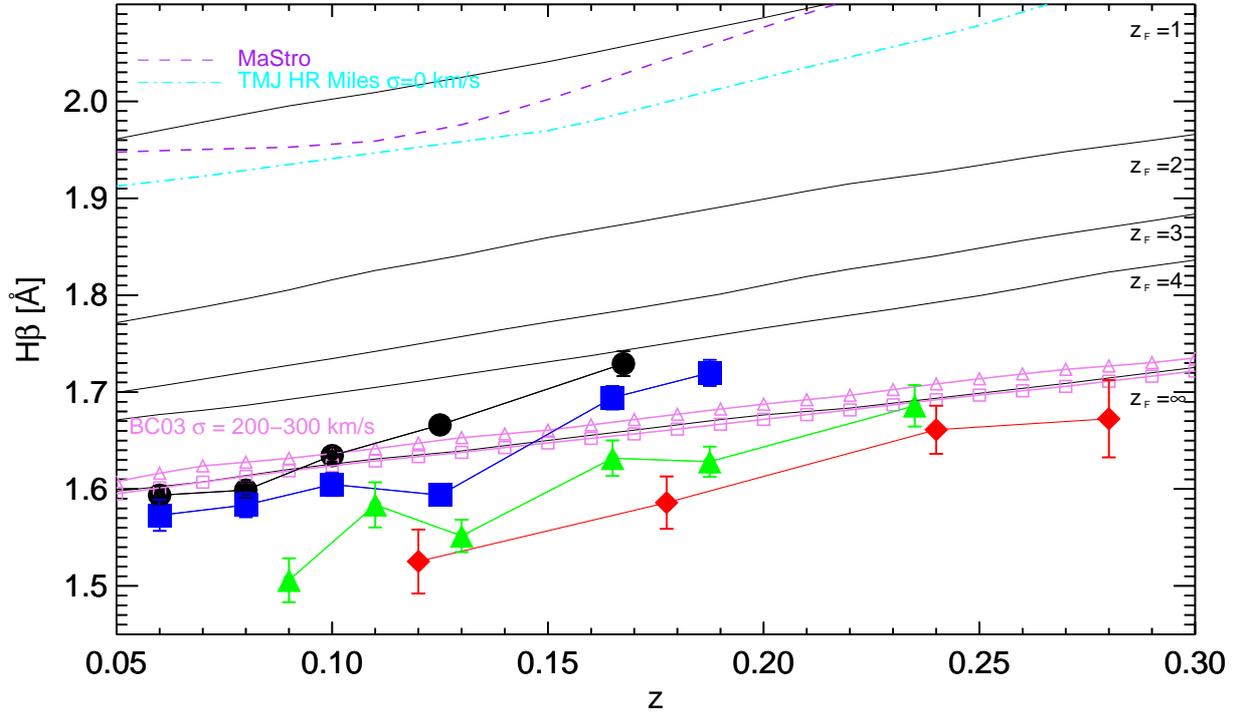}  
\caption{{Comparison between the observed H$\beta$-z relations (circles, squares, triangles and diamonds for BinI, BinII, BinIII and BinIV respectively) and different models. The black solid lines are the BC03 models with $\sigma =230$ km s$^{-1}$ $Z/Z_{\odot}=1.2$ and different formation redshifts ($z_F =1$, $2$, $3$, $4$ and $\infty$). For $z_F=\infty$ we show also the models with $\sigma=200$ and $300$ km s$^{-1}$, open triangles and open squares respectively. The MaStro and TMJ models are shown for $z_F=1$ (dashed and dash and dot lines, respectively). (A color version of this figure is available in the online journal.)} }
\label{fig:HbetaconModelli}
\end{figure*}
 \subsubsection{Effects of the metallicity}

The models are also used to establish the effect in the H$\beta$ index due to the stellar metallicity.
While there is no significant difference in the median metallicity of the different
mass subsamples, each mass bin instead shows a slight decrease of the median metallicity with redshift. 

We create a library of H$\beta$ indices as a function of age using stellar population models with different metallicity: $1 $, $ 1.4 $ and $1.5$ $ Z / Z_ {\odot} $. 
The study of this H$\beta$ synthetic indices shows that even an extreme variation of metallicity, between $Z/Z_ {\odot} =1$ to $Z/Z_ {\odot} =1.4,1.5$, implies a H$\beta$ variation of only $\sim 4-5\%$ on average, always lower than H$\beta$ trend observed.
Moreover, to further check the metallicity effect, not only in the models but also on our data, we create a control sample of  galaxies with a narrow metallicity range (e.g., $\Delta(Z/Z_ {\odot}) =0.5$ for the BinI).
Even in this subsample we find a steady trend of growth in the H$\beta-z$ relations.

From these analyses, we find that the increase of the H$\beta$ equivalent width with redshift (at fixed mass) and decrease with mass (at fixed redshift) of passive galaxies cannot be only due to $\sigma$ or $Z/Z_{\odot}$ variation, and hence can be mostly explained by an age-evolution effect, in particular with the age of the stellar population decreasing with increasing redshift, and increasing with increasing mass.


\subsection{H$\beta$-age: Comparison with SPS models} \label{compModels}

To interpret the evolution of H$\beta$ with redshift and its dependence on mass in the context of passive galaxies evolution, we proceed with a direct 	comparison between the observed data and the index values calculated on the SED provided by SPS models, hereafter theoretical H$\beta$. Both the observed and theoretical measurements are obtained by the same method, described in Section \ref{index}.
The theoretical H$\beta$ curves as a function of age are transformed into H$\beta$-z relations assuming a $\Lambda$CDM cosmology (see Section \ref{introduzione} for the parameters used), and a formation redshift.
  For our studies, we chose to probe different redshifts of formation for our galaxies, namely $z_F=1, 2, 3,4$ and $\infty$. For this comparison we use two different stellar populations synthesis models, BC03 and MaStro
and the theoretical models of H$\beta$ Lick indexes of \cite{Thomas2011} (hereafter TMJ), calculated by theoretically manipulating the index response functions.
For BC03 and MaStro models, we adopt a Chabrier IMF instead the TMJ models have a Salpeter IMF. 
However, using BC03 models, we verify that a IMF variation does not affect significantly the H$\beta$ value. In particular, in the case of solar metallicity, the choice of a Salpeter IMF instead of a Chabrier causes an index decrease of$<1$\%.
Furthermore, in the BC03 and MaStro models we use a delayed exponentially SFH, in particular we use models with $\tau = 0.2$ Gyrs which correspond a the $\tau$ median value estimate for the our sample (see Section \ref{selection} ). 
We investigate also the possible variation of H$\beta$ value with different $\tau$ values. We compare SPS models with $\tau=0.05$ and $\tau=0.2$ Gyrs, and we find that for SPS older than $6$ Gyr the H$\beta$ Lick index decreases of $\leq 1$ \% in the SPS with $\tau=0.05$ Gyrs to respect at SPS with $\tau=0.2$ Gyrs.
Since the median metallicity of the sample is $Z/Z_{\odot}\sim1.2$, for the comparison with the data we interpolated quadratically the H$\beta$ values of all models in the three metallicities provided, in order to have an estimate of the index at $Z/Z_{\odot}=1.2$. The BC03 and MaStro models are convolved with a velocity dispersion $\sigma=230$ km s$^{-1}$, comparable to the median value of our sample.
In Fig. \ref{fig:HbetaconModelli}, we show as an example the observed H$\beta-$z relations and the theoretical curves (BC03 models) for different formation redshifts. For MaStro and TMJ models we find similar results, as shown in the case of formation redshift $z_F = 1$.

From the comparison with theoretical models, 
as can be clearly seen from Fig. \ref{fig:HbetaconModelli}, 
we find that they do not reproduce the observed H$\beta$ median relations, since, 
the models predict higher H$\beta$ values than the median H$\beta$ measured on our samples, so that in most cases the observed values would require an age greater than the age of the Universe.
For the most massive sample the minimum shift between data and models is of the order of $\Delta$ H$\beta = 0.1$.

Since the models are built with the same metallicity and velocity dispersion found in the data, we exclude that this inconsistency is due to one of these two parameters.
We verify that also by imposing an extreme velocity dispersion for our sample ($\sigma = 300$ km s$^{-1}$) the models do not reproduce the observations. 
Regarding the metallicity, we instead find that the H$\beta-$z relations can be reproduced only by considering models with metallicity approximately $Z/Z_{\odot}=2$, that are inconsistent with the metallicity estimated for this sample. 

Furthermore, comparing the metallicity estimated by \cite{Thomas2010} for a morphological selected sample of ETGs at $0.05\leq $z$\leq 0.06$, we notice that for a subsample with velocity dispersion similar to our values they estimate a slightly higher metallicity than those considered in this work \citep{Gallazzi2005}. 
In particular, for $\sigma = 207$ km s$^{-1}$ (median value of the less massive sample, BinI) \cite{Thomas2010} find a metallicity $Z/Z_{\odot} \sim 1.5 $, and for the median value of the total sample, $\sigma = 230$ km s$^{-1}$, they estimate a metallicity of the order of $Z/Z_{\odot} \sim 1.6 $ \citep[for more details, refer to][]{Thomas2010}. 
These higher metallicity values are also confirmed in a more recent work by \cite{Citro2015}. In this study the authors find a median $Z/Z_{\odot} \sim 1.450 \pm 0.075 $.
Even these higher values are not enough to completely reconcile models with data.

\section{Investigating the origin of the offset}\label{discussion}
In this section, we further investigate possible causes of the offset between the model predictions and the measures of the H$\beta$ Lick index showed in Section \ref{compModels}.
In particular, we explore the possibility that our sample, despite the selection criteria (see Section \ref{selection}), is still contaminated by an emission-line component.
{This potential emission component would contaminate the measurement of the H$\beta$ index by filling the line and producing a less intense absorption feature, consistent with the recent results showed by \cite{ServenWorthey2010} for a SDSS low redshift ($0.06<z<0.08$) galaxy sample.}


\begin{figure*}
\centering
\includegraphics[angle=-90, width=\hsize, trim = 0mm 0mm 0mm 10mm, clip=true]{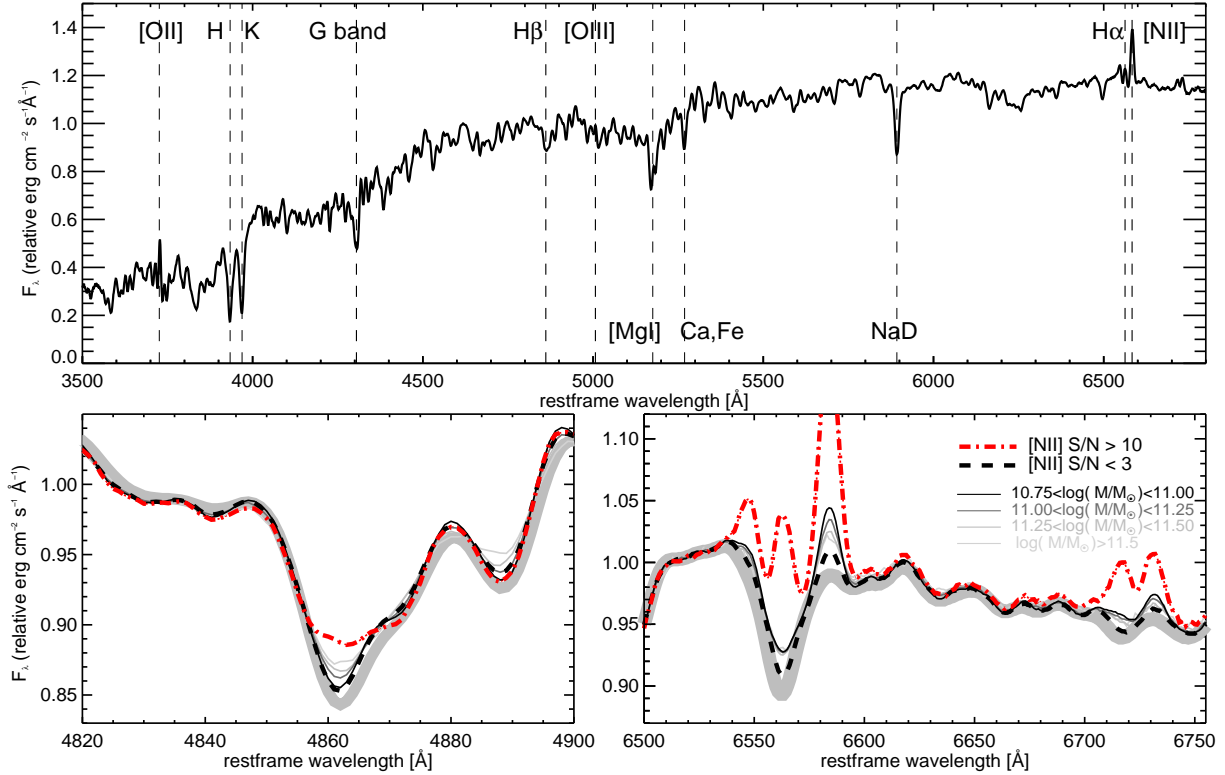}
\caption{{Top panel: stacked spectrum of the sample with [NII]$\lambda$6584 S/N$>$10 ($1145$ galaxies), it shows some evident emission lines: [OII], [OIII], H$\alpha$ and [NII]$\lambda$6584. Bottom panels: comparison 
between the [NII]$\lambda$6584 S/N$>$10 stacked spectrum (red dash dot lines), [NII]$\lambda$6584 S/N$<$3 stacked spectra (black dashed lines) and the stacked spectra of the original sample in different mass subsamples (from black to light gray for BinI, BinII, BinIII and BinIV respectively). In the H$\beta$ region (left panel) are not evident emission lines, instead, in the H$\alpha$ region there are evident emissions ([NII]$\lambda$6548, H$\alpha$, [NII]$\lambda$6584, [SII]$\lambda$6717 and [SII]$\lambda$6731). }The gray area is the SSP solar model with age$=10$ Gyrs, $\sigma=250$ km s$^{-1}$ calculated by BC03 model. (A color version of this figure is available in the online journal.)}
\label{fig:ave_spectraNIImaggiore10}
\end{figure*}

\subsection{Improving sample selection: the [NII] emission-line contamination}\label{nii_emission}
%

As discussed in Section \ref{sample}, we select only the galaxies that not showing significant emission lines in H$\alpha$ and [OII] (EW$<$-5~\AA). 
We check that also cutting the sample by adopting a more stringent H$\alpha$ limit (S/N$<3$), the offset between the H$\beta$ lick index in our data and models is not removed.
However, a more detailed analysis of the stacked spectra shows the presence of a weak [NII] emission at $6584$~\AA\ (see Fig. \ref{spettri}).
Since the first ionization potential of the nitrogen is very similar to that of hydrogen, the presence of an emission line in [NII]$\lambda$6584 may indicate the presence of an emission component also in the H lines.
In order to study the possible emission line contamination in our sample, we use the EW([NII]$\lambda$6584) values provided by the MPA-JHU group.
From this analysis, about $43 \%$ of the total sample have an equivalent width EW([NII]$\lambda$6584) with S/N$>$3, consistent with the clearly visible [NII]$\lambda$6584 emission line in the median spectra (see Fig. \ref{spettri} left panel).
The median and dispersion\footnote{The dispersion is evaluated whit the $median$ $absolute$ $deviation$, $MAD$.} equivalent width of the total sample are EW([NII]$\lambda$6584)=-0.56$\pm$0.58~\AA, in agreement with the equivalent width measured on the median stacked spectrum (EW([NII]$\lambda$6584)=-0.68~\AA).

\begin{figure}
\centering
\includegraphics[angle=-90, width=\hsize ]{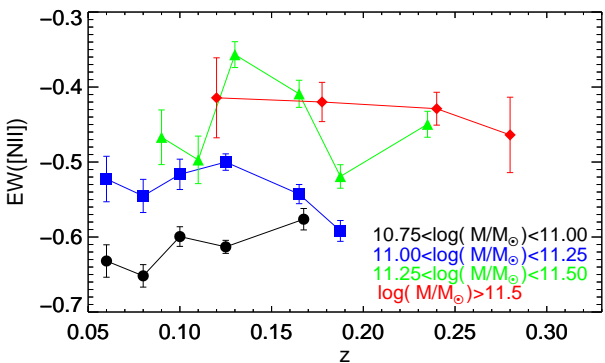}
\caption{{[NII]$\lambda$6584-redshift relations for the four mass subsamples (circles, squares, triangles and diamonds for BinI, BinII, BinIII and BinIV respectively). (A color version of this figure is available in the online journal.)}}
\label{fig:NIImediano4Bin}
\end{figure}

Moreover, by analyzing separately each mass subsample, we identify a relationship of the [NII]$\lambda$6584 emission line as a function of stellar mass. In Fig. \ref{fig:ave_spectraNIImaggiore10}, bottom right panel, we show the stacked spectra of the four mass subsamples, from which it is evident that the [NII]$\lambda$6584 emission line increases with decreasing of the median mass.
This evidence is confirmed also by analyzing the EW([NII]$\lambda$6584)-z relations: we note that the EW([NII]$\lambda$6584) median values are almost constant with redshift, and they decrease with increasing mass (see Fig. \ref{fig:NIImediano4Bin}).

%
%
%

To verify if the presence of a [NII]$\lambda$6584 emission is related to the presence of H$\beta$ emission line, we analyze the galaxies most affected by this contamination.
We select the galaxies with a [NII]$\lambda$6584 clearly detected, i.e. with S/N([NII]$\lambda$6584)$>$10; in this way, we select $1145$ galaxies\footnote{In this work, we decide to cut the sample above this limit ( S/N([NII]$\lambda$6584)$>$10 ) because enabled us to select only those objects in which the [NII] line is detected effectively while maintaining a good statistic (1145 galaxies).} (5\% of the total sample). Figure \ref{fig:ave_spectraNIImaggiore10} shows the median spectrum relative to this subsample. It can be seen that, differently from the stacked spectrum of the global sample (see Fig. \ref{spettri}), in this spectrum, together with a strong [NII]$\lambda$6584, there are clearly detectable several emission lines, i.e. [OII], [OIII]$\lambda$5007, [NII]$\lambda$6548, 
H$\alpha$,  [SII]$\lambda$6717 and [SII]$\lambda$6731 (Fig. \ref{fig:ave_spectraNIImaggiore10}). Nevertheless, also in this case, in the H$\beta$ region, there is not a clear emission component (Fig. \ref{fig:ave_spectraNIImaggiore10} bottom left panel). The H$\beta$ absorption line only shows a particular shape, different from that expected for a simple absorption, probably caused by an overlapping with a weak emission line contamination.

Following this result, we decide to apply a further cut to our sample, by selecting only the galaxies with a [NII]$\lambda$6584 line detected with S/N([NII]$\lambda$6584)$<$3; this sample is composed by $13626$ galaxies ($\sim$57\% of the total sample). We find that, even for this sample there is a residual [NII]$\lambda$6584 line, with a median and dispersion (MAD) EW([NII]$\lambda$6584)$\sim -0.26\pm 0.27$ \AA.
 The residual emission can be seen from the comparison between the stacked spectrum of this sample and a pure absorption SPS model (see Fig.\ref{fig:ave_spectraNIImaggiore10} bottom panels).

\begin{figure*}
\centering
\includegraphics[angle=-90, width=\hsize]{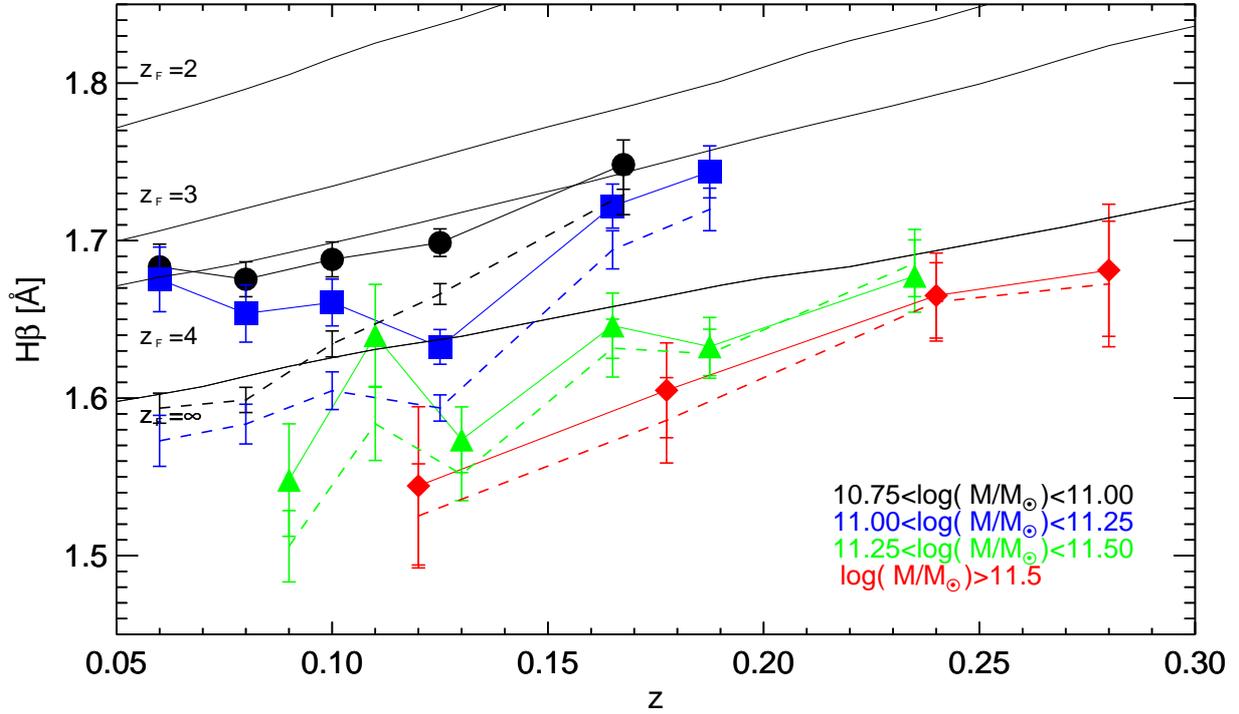}
\caption{{Comparison between H$\beta$-z relations (circles, squares, triangles and diamonds for BinI, BinII, BinIII and BinIV respectively) of the total sample (dotted curves) and the sample with EW([NII]$\lambda$6584) S/N$<$3 (continuous curves ). (A color version of this figure is available in the online journal.)}}
\label{fig:HbetaConfrontoNIIminore3}
\end{figure*}

On this new subsample with S/N([NII]$\lambda$6584)$<$3, we calculate also the median H$\beta$-z relations at different mass, the results are shown in Fig. \ref{fig:HbetaConfrontoNIIminore3}. 
We note that the cut in [NII]$\lambda$6584 produces a general increase in median values of the median H$\beta$ index (on average to $\sim 0.05$ \AA\ for the BinI and BinII, $\sim 0.02$ for the BinIII and $\sim 0.01$ for the BinIV ). The offset found between data and observations of the total sample is reduced.
However, there is still a significant disagreement between data and theoretical H$\beta $-z curves, especially for the more massive sample (Bin IV).
However, despite this discrepancy, we find similar H$\beta$-z and H$\beta$-Mass relationships as seen in the total sample, but with slightly different slopes and normalizations. These variations make impossible to use the median H$\beta$ values for an accurate estimate of the absolute and differential age evolution of the stellar populations with the redshift. Consequently, it is not possible to place constraints on the redshift of formation of the galaxies and even the use of the H$\beta$-z relation in the method of the cosmic chronometers appears inadequate.

\subsection{Methods to remove nebular emission}
The SDSS galaxies spectra are extensively used for many purposes, such as stellar mass and star formation history estimates \citep{Kauffmann2003}, age and metallicity estimates in the local universe \citep{Gallazzi2005}, environment effects in the galaxy formation \citep{Thomas2010}, and many other.
In these previous works, before interpreting the observed spectra, and therefore before using the stellar absorption-line indices, standard procedures are followed to remove the contamination by nebular emission lines. In this section we try to apply these cleaning methods both in the individual and in the stacked galaxies spectra. 

The method proposed by \cite{Tremonti2004}, for example, makes a non-negative least-squares fit of the emission line free-observed regions of the spectrum, using a spectral library built using BC03 models. Then, the fitted spectrum is subtracted from the observed spectrum, and the residuals can be fitted to Gaussian broadened emission lines templates. 
Finally, the fitted emission lines are subtracted from the original observed spectrum to produce a ``pure'' absorption line spectrum \citep[for further details, we refer to][]{Tremonti2004}.
To verify if this method of decontamination is actually able to producing a clean sample in the H$\beta$ region, we use H$\beta$ values
measured by the MPA-JHU DR7 group, which are available both for the original observed spectra that for the "pure" absorption line spectra.
We select a sample consisting on galaxies without emission lines, i.e. with no correction for emission lines based on the method of \cite{ Tremonti2004}, which contains 19904 objects.
We measure the median H$\beta$-z relations also for this subsample, with the same procedure described in Section \ref{results}.
The H$\beta$-z relations obtained are consistent with those found for the subsample with S/N([NII]$\lambda$6584)$<$3. 
This suggests that, using objects with no emission lines detected by standard correction method, the offset between the observed H$\beta$ values and the predicted ones is not totally removed. 
\\
In addition, we test the presence of weak H$\beta$ emission line directly in the stacked spectra. We split our sample in terms of stellar mass and redshift as we done in Section \ref{results} for the H$\beta$ values, and then we built the median spectra for each mass and redshift bin. In order to separate the stellar continuum from the nebular emission component, we decide to using a combination of the publicly available IDL codes: penalized pixel-fitting (pPXF), devoloped by \citep{CappellariEmsellem2004} and gas and absorption line fitting (GANDALF) write by \citep{Sarzi2006}. 
We measure the line-of-sight velocity distribution (LOSVD) by using pPXF code. Then, we perform the GANDALF analysis to convolve a set of input synthetic spectra with the previously kinematic and to fit the observed stacked spectrum simultaneously with the models and a Gaussian emission lines template. The result is a superposition of an optimal combination of the SSP templates with a set of Gaussians that represent the emission lines.
Through the subtraction of the emission line spectrum from the observed one, we get the clean absorption line spectrum free from emission line contamination. First, we analyze the obtained H$\beta$ emission lines and then we explore the impact of this cleaning method in the observed spectra.
In order to test the robustness of the emission lines extraction, we repeat the methodology by using three different spectral library: two based on the BC03 SSP \footnote{The BC03 and MaStro stellar population templates have a Chabrier IMF and metallicity $Z/Z_{\sun} = 0.2,0.4,1,2.5$ and $Z/Z_{\sun} = 0.5, 1,2$, respectively.} with different age $0.01\leq t \leq 14$ Gyr (hereafter BC03$\_14$Gyr) and a more extended age $0.01\leq t \leq 20$ Gyr (hereafter BC03$\_20$Gyr); and the third library built with MaStro models with and age $0.01\leq t \leq 14$ Gyr. 
We find that, the recovered H$\beta $ emission lines are model dependent: the equivalent width is systematically higher for 
the BC03$\_20$Gyr library than the BC03$\_14$Gyr ($\Delta$EW$\sim 0.02$ \AA ) and the differences increase if we compare BC03 with MaStro models, $\Delta$EW $\sim 0.2$ \AA\ . The median  equivalent width are : EW$\sim -0.09 \pm 0.005$, $\sim -0.07 \pm 0.01$ and $\sim -0.27 \pm 0.01$ \AA\, for BC03$\_14$Gyr, BC03$\_20$Gyr and MaStro library, respectively. This result is consistent with the study showed in \cite{Singh2013}; their figure $4$ shows that at low H$\beta$ fluxes there are differences in the flux extraction by using GANDALF code with SSP instead of the stellar templates. Therefore, the correction for very low emission in the absorption H$\beta$ line depends on the templates library used in the continuum fit. Furthermore, we find that, in all the three libraries, the H$\beta$ emission lines are detected with a very low confidence level, the amplitude over noise is always A/N$\leq 3$ (the noise is defined as the dispersion of fluctuations in the fit residuals).

{Finally, we test the more recent emission correction for the hydrogen features proposed by \cite{ServenWorthey2010}. \cite{ServenWorthey2010} derived emission corrections of the Balmer series Lick indices for a SDSS quiescent galaxy spectral sample (\citealp{Graves2007}) by comparing the H$\alpha$-Mg b diagram from the SDSS stacked spectra with the measurements obtained for 13 Virgo galaxies. By using the same prescriptions showed in \cite{ServenWorthey2010}, we re-calibrate the H$\alpha$-Mg b diagram and then the H$\beta$ emission correction in our stacked spectra. We find that for the more massive sample (BinIV), the mean Mg b value is $4.23 \pm 0.15$ \AA\ and the mean H$\beta$ emission correction factor is on the order of $-0.17 \pm 0.08$ \AA\ , that it is in good agreement with the emission values obtained with the GANDALF method mentioned before. 
However, we stress that, also in this case, the method is model-dependent, since the continuum correction was determined by using the \cite{Worthey1994} and \cite{Trager1998} models (see Section 2 of \citealp{ServenWorthey2010}). Moreover, rescaling the H$\alpha$-Mg b relation from the SDSS values to the Virgo data could be age-dependent if the two samples have different mean ages. We conclude that for our sample, also in the stacked spectra, is very difficult measure a well detected and model-age-indipendent H$\beta$ emission line and than correct our observed spectra in order to obtain a "pure" absorption H$\beta$ line.}

\subsection{H$\beta$ self-diluted emission line}
\label{hiddenemission}
\begin{figure*}
\centering
\includegraphics[angle=-90, width=\hsize, trim = 0mm 0mm -10mm 0mm, clip=true]{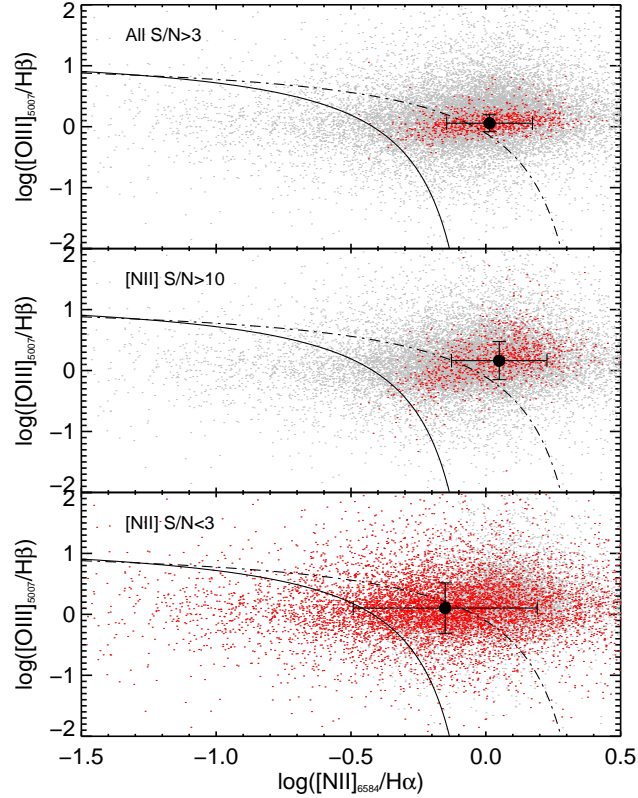}
\caption{Distribution of the galaxies in the BPT diagram for the sample with all EW S/N$>$3, EW([NII]$\lambda$6584) 
S/N$>$10 and EW([NII]$\lambda$6584) S/N$<$3, respectively from the top downwards. The gray cloud is the total sample.
The dashed curve is the theoretical demarcation of \citet{Kewley2001} , that separates star-forming galaxies and composites from AGN. 
The solid curve indicate the empirical division between pure star-forming galaxies from composite and AGN of \citet{Kauffmann2003}.
The median value and dispersion of the distribution are shown with the black symbols, in each subsample. {(A color version of this figure is available in the online journal.)}}
\label{fig:BPT}
\end{figure*}

In order to have a qualitative estimate of the hidden EW(H$\beta$) emission related to the observed [NII]$\lambda$6584 emission line, we study the emission line ratios of our spectra in the diagnostic diagrams.
We adopt the classical \citep{Baldwin1981} diagram, hereafter BPT diagram.
The equivalent width measurements for [OIII], H$\beta$, H$\alpha$ and [NII]$\lambda$6584 are taken by the MPA-JHU group analysis.
In Fig. \ref{fig:BPT} we show the BPT diagrams for the subsample with all EWs detected with S/N$>$3, the subsample with EW([NII]$\lambda$6584) with S/N$>$10, and the subsample with EW([NII]$\lambda$6584) with S/N$<$3.
In the first two cases, we find that the majority of the objects are located in the LINERs region \citep[Low Ionization Nuclear Emission line Regions,][]{Heckman1980}. 
For the first and the second sample, the median and dispersion for
$\log(EW ([NII])/EW(H\alpha))$ are $-0.014\pm 0.159$ and $0.050\pm 0.177$, and for $\log(EW([OIII])/EW(H\beta))$ are $0.058\pm  0.139$ and $0.162\pm 0.312$, respectively.
In the last subsample, the dispersion in the data is larger because of the lower signal to noise ratio (see Fig. \ref{fig:BPT}).
In this last case, the median value of the sample is located in the composite region ($\log(EW ([NII])/EW(H\alpha)) \sim -0.150 \pm 0.341$  and $\log(EW([OIII])/EW(H\beta)) \sim 0.103 \pm 0.413$).

By using the median EW([NII]$\lambda$6584)/EW(H$\alpha$) values and the median EW([NII]$\lambda$6584) we estimate qualitatively the EW(H$\alpha$) contribution.
Then, we estimate the EW(H$\beta$) by assuming no absorption by dust, and hence an emission line ratio, H$\alpha /$ H$\beta$ $\sim 2.86$ \citep{Osterbrock1989}, for electron density of $n=100$ $cm^{-3}$ and electron temperature $T_e =10^4 K$).

In the sample with EW([NII]$\lambda$6584) with S/N$<$3 the EW([NII]$\lambda$6584) is $\simeq$-0.26$\pm$0.27 \AA. For this sample we expect an EW(H$\beta$) emission in the range $-0.28$ to $-0.06$ \AA\ (by considering the EW([NII]$\lambda$6584)/EW(H$\alpha$) dispersion), or weaker in case of dust extinction.

The same conclusions can be drawn for the sample with the larger [NII]$\lambda$6584 emission lines (S/N([NII]$\lambda$6584)$>$10). 
In this case, the median EW([NII]$\lambda$6584) is -1.99$\pm$0.58~\AA\ , and the interval of variability of $log(EW([NII])/EW(H\alpha)$ is smaller, $\sim 0.050 \pm 0.177$. 
Then, we expect an H$\beta$ emission line equivalent width between $-0.9$ and $-0.4$~\AA\ in case of no absorption by dust, or weaker otherwise.


In order to verify whether the H$\beta$ emission line values, expected in the two different samples, S/N([NII]$\lambda$6584)$<$3 and $>$10,
can be effectively detected in the observed spectra, we perform some simulations, described in the following section.

\subsection{H$\beta$ emission line simulations}
\begin{figure*}
\centering
\includegraphics[angle=-90, width=\hsize, trim = 0mm 0mm -10mm 0mm, clip=true]{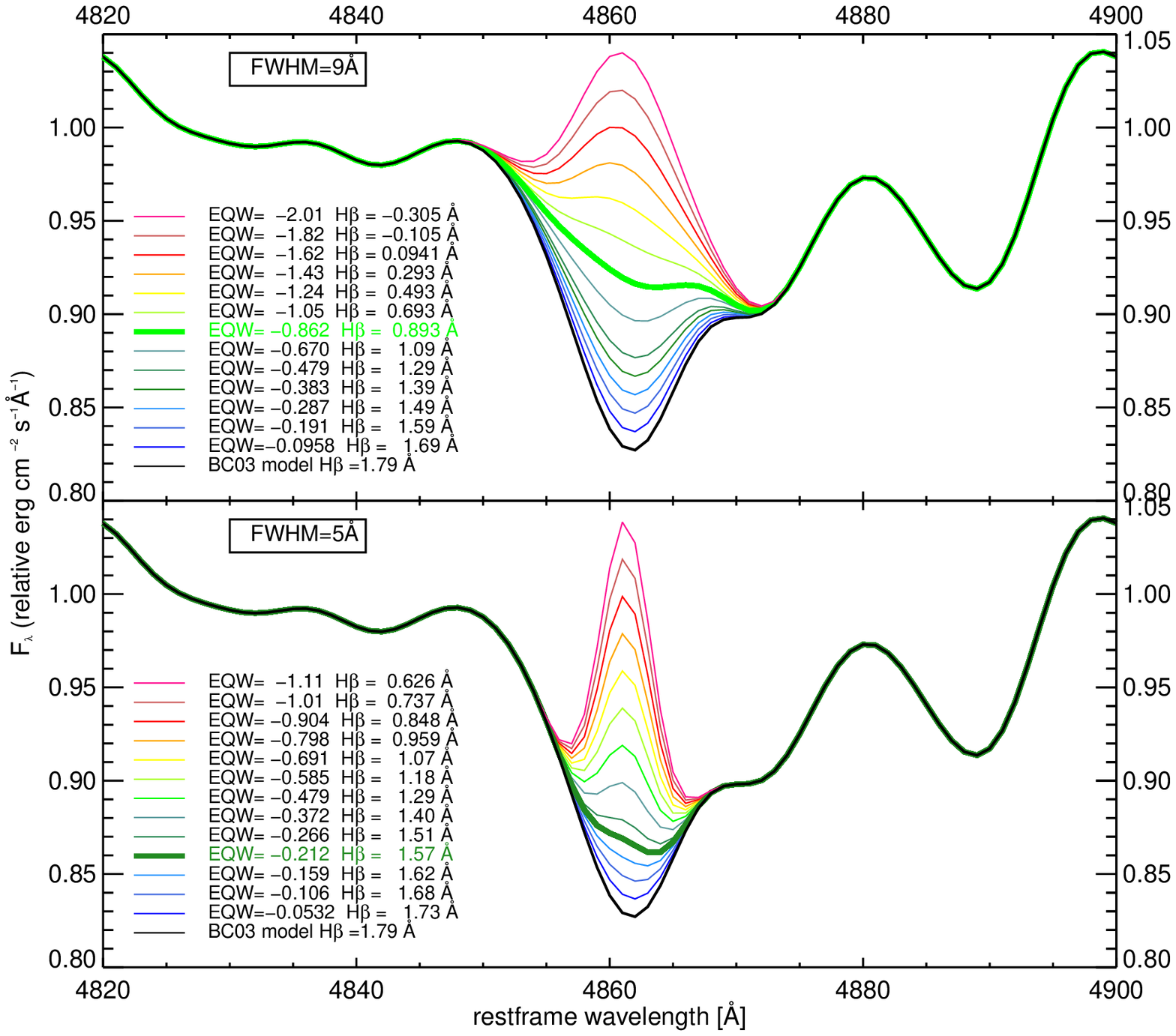}
\caption{H$\beta$ emission line simulations. We show the emission line effect within the H$\beta$ absorption line (Simul A in the lower panel and Simul B in the upper panel). {The different EW contaminations are plotted according with the labels (with different colors in the online journal).} In both figures we can identify a threshold limit above which it is possible to detect an emission contamination (thick curves). The absorption model is a SSP with $Z/Z_{\odot}=1$, age$=10$ Gyrs and $\sigma=200$ km s$^{-1}$. }
\label{fig:emissionLine} 
\end{figure*}

Starting from the theoretical SED (in which only the absorption component is taken into account) of SPS models (BC03), we simulate the presence of an emission line by adding a Gaussian component with variable amplitude and FWHM. 

Since different physical processes can give rise to emission lines with different FWHM, we perform two different simulations:
\begin{itemize}
\item {\bf SIMUL A}, lines with FWHM=5 \AA, mainly due to star formation. The analysis of a spectrum of a typical star-forming galaxy from SDSS has shown a typical FWHM of the emission lines of $ \sim 5$ \AA.
\item {\bf SIMUL B}, lines with FWHM=9 \AA. 
Since, in the case of emission lines due to different processes than star formation, such as AGN activity, the lines are broader than the ones found for star-forming galaxies.
\end{itemize}

By changing the FWHM and the EW line in theoretical emission lines, we study the variation of the absorption line profile with the contamination of this different emissions. The results of the simulations (for the case of a SSP SED with age=10 Gyr) are shown in Fig \ref{fig:emissionLine}.
In both simulations (A and B) it is possible to identify a threshold of emission below which it is not possible to detect the H$\beta$ emission component on the absorption line. 
We can appreciate the distortion of the H$\beta$ line due to the introduction of an emission line, with increasing EW, only after EW $\leq -0.2$ \AA\ or EW $\leq -0.9$ \AA\ for SIMUL A or B, respectively. Emissions smaller than these values are not able to change the shape of the line, but however, they cause a intensity decrease in the absorption line, producing a drastic variation in the H$\beta$ lick index measure, with $\Delta($H$\beta)_{max} \sim 0.2$ \AA\ or $\Delta($H$\beta)_{max} \sim 0.9$ \AA\ in SIMUL A and B, respectively.
From this analysis, we can see that, for galaxies spectra with a resolution of the order of $3$ \AA\ FWHM, there is a threshold detection limit of the H$\beta$ emission line. As seen in previous section, the presence of [NII]$\lambda$6584 emission in our samples may indicate the presence of a weak H$\beta$ emission line with EW  $\geq -0.28$ \AA\  for the sample whit S/N([NII]$\lambda$6584)$<$3 and EW $\geq -0.9$ \AA\  for the sample whit S/N([NII]$\lambda$6584)$>$10.
These values are near to the threshold limit identified by simulations, therefore consistent with not being detectable in emission in our spectra. However, this potential hidden emission line can explain the detected offset between observational data and theoretical models discussed in the Section \ref{compModels}. 

Performing the same simulations for SSP models with different ages (5 and 13 Gyr), we find that the detection limit for the H$\beta$ emission line does not change significantly.

\subsection{Ionization sources}
We find that even the less contaminated passive galaxies (sample with EW([NII]$\lambda$6584) S/N$<$3) may present an emission line contamination in the H$\beta$ absorption line. 
As discussed in Section \ref{hiddenemission} using the BPT diagrams, we found that by selecting samples with  higher S/N (sample with all lines S/N$>$3 and sample with EW([NII]$\lambda$6584) 
S/N$>$10), our galaxies show tipically LINERs-like rather than star-formation emission line ratios.
These results are consistent with several imaging and spectroscopic observations \citep{Phillips1986,Goudfrooij1999,Sarzi2006,Annibali2010,Yan2012}, which show a strong indication of the presence of warm gas (typically with $T\sim 10^{4}$ K) in ETGs, with line ratios typically classified as LINERs.
Despite the number of studies, two fundamental questions remain open: what is the origin of this ISM in the ETGs and what is the physical mechanism that determines its ionization?\\
Studies carried out on the kinematics of the gas, in most cases, show a misalignment between gas and stellar component, suggesting an external origin of the gas \citep{Caon2000}. This observational evidence is in agreement with recent estimates of the gas emission metallicity, which seems to indicate an external source \citep{Annibali2010}.

However, as reported in \cite{Sarzi2006}, the angular momenta measured do not seem to be consistent with a purely external origin. 

The nature of the mechanism of ionization in LINERs is still not clear, but currently, the most accredited are photo-ionization by AGN activity \citep[e.g.][]{Ho1999,Ho2000,Kewley2006}, photo-ionization by post-asymptotic giant branch (post-AGB) stars \citep{Binette1994} and fast shocks \citep[e.g.][]{Dopita1995}.
The AGN activity, in fact, produces a source of energetic photons (X-rays and UV) able to ionize the interstellar medium; this is confirmed by some observed bubbles in LINERs galaxies \citep{Baldi2009}.
For some objects, the presence of AGN activity was also confirmed by observations of the radio core and X-rays point sources in their centers and UV variability \citep[e.g.][]{Ho2008}. 
Nevertheless, recent works have shown that in many cases the gas emissions are not only neighboring to the center, but an extended emission component also exists \citep{Sarzi2006,Annibali2010,Yan2012,Singh2013}.
For this reason, the AGNs are not considered as the main cause of the LINERs-like emission lines in ETGs in favor of spatially extended ionizing sources, that in some cases follow the profile of stellar density \citep{Yan2012}.
These properties are found in another source proposed for the first time by \cite{Binette1994}: the post-AGB stars in the old stellar populations. These stars, after the AGB phase, are often surrounded by material ejected during their thermal pulses. After that, this material is dispersed in the ISM, and the AGB stars, having very high temperatures ($\sim 10^5 K$), are capable of producing a diffuse radiation field. 
Recent spatially resolved spectroscopy observations obtained with a prototype of MANGA instrument, confirm the presence of an extended LINERs-like emission associated with spectral features of old and metal rich stars, see \cite{Belfiore2015}.
This scenario is also supported by the fact that post-AGBs are able to reproduce the variation of the parameter of ionization (defined as the ratio of the ionizing photon flux density to the electron density) with the radius of the galaxy \citep{Yan2012}.
As expected in \cite{Binette1994} and \cite{CidFernandes2009}, the post-AGBs are able to produce weak emission lines: H$\alpha \sim - 0.6$, $-1.7$ \AA\, for a stellar population of $8$ and $13$ Gyrs respectively. 
This is very interesting because if this emission is expressed in terms of H$\beta$ emissions, similar to what was seen in the Section \ref{hiddenemission}, we find a measure of 
EW(H$\beta$)$\sim-0.2$, $-0.6$ \AA\ emission line perfectly consistent with what is expected for our sample,  especially with regard to the subsample with EW([NII]$\lambda$6584) S/N$<$3 \AA\ .
However, as shown in \cite{Yan2012}, the post-AGB stars, while representing the ideal candidate for the creation of an photo-ionizing source, according to the most recent models, produce an ionization parameter too small compared to what is required by the observations. 
It should be noted that the current knowledge of the number density of post-AGB stars is still very uncertain, and therefore the question if these stars may be the main cause of emissions in older populations remains open.
Finally, the emission lines with LINERs-like ratios can be produced by fast shock. However, this mechanism can not be mainly responsible for the emission in ETGs, since, as suggested in \cite{Yan2012}, it provides emission lines with similar velocities between the different elements, contrary to what emerges from the recent estimates of velocity in different lines.

The data analyzed here do not give a direct indication of how to disentangle the three possible photo-ionization source candidates. However the observed anti-correlation between the [NII] emission and the mass (see Sec. \ref{nii_emission}, Fig. \ref{fig:NIImediano4Bin}) could disfavor the AGN hypothesis as a photo-ionization source, since its presence should instead be correlated with the galaxy mass \citep[see][]{Best2005,Brusa2009}.

Finally, it is particularly important that the detection of emission lines in our sample seems to suggest a more frequent presence of ionized gas in these galaxies. This observation has significant effects on knowledge of this particular class of galaxies. The ETGs are not the simplest systems, but they are more complex in which only a knowledge of both the stellar contribution and of the ISM (and of the link between the two components) can shed light on their formation history and evolution.


\section{Conclusions}  \label{conclusioni}

In this paper, we explore the properties of the H$\beta$ Lick index of massive and passive ETGs, to estimate its robustness as an age indicator, being the Lick index more sensitive to the stellar population age and less affected by stellar metallicity, as suggested since the pioneering work of \cite{Worthey1994}. The aim of this work is to establish its reliability as ``cosmic chronometer'' to trace the age evolution of the Universe as a function of redshift and to provide new constraints on the age of formation and evolution of galaxies stellar populations, eventually allowing measurement of the Hubble parameter $H(z)$ through the ``cosmic chronometers'' approach \citep{Jimenez2002,Moresco2012a,Moresco2015,Moresco2016a}.

Using photometric and spectroscopic information, we select the most massive, passive and red ETGs in the SDSS-DR6 survey.
The final sample consists of about $23914$ galaxies with stellar mass $\log(M/M_{\odot}> 10.75)$ in the redshift range $0.05\leq z \leq 0.3$.
We divide the sample into four mass subsamples ($\Delta \log(M/M_{\odot}) =0.25$) in order to avoid possible biases due to mass downsizing effect, and obtained four homogeneous subsamples in redshift of formation.
All the samples are further divided into redshift bins and the spectra analyzed to measure the H$\beta$ Lick index, obtaining a median $H\beta-z$ relation in each mass subsample.

The main results of this analysis may be summarised as follows.

\begin{itemize}

\item Despite the rather strict selection criteria, all the spectra present characteristic features of a passive population, we find in the median stacked spectra of each mass bin clear evidence of a weak [NII] emission line, with equivalent width of the order of $-0.6$ \AA, which can be interpreted as a hint of the presence of ionized gas. This residual emission line contamination may have a significant impact on the H$\beta$ Lick index measures. To address this issue, we split our samples on the base of
the [NII] line, having a purer sample of 13626 galaxies selected with SN([NII])$<$3, and a more contaminated sample with SN([NII])$>$10. The analysis of these two subsamples
confirmed the first hypothesis of a possible hidden contamination, with the sample with SN([NII])$>$10 presenting clear emission lines also in H$\alpha$,[OIII] and H$\beta$. We also find that the amplitude of the [NII] emission line anti-correlates with stellar mass.

\item The analysis of all mass subsamples reveals a clear evolution of the H$\beta$ Lick index as a function of redshift, with the index decreasing with cosmic time. These trends are qualitatively consistent with a passive evolution, and are proven to be independent on many effects that can affect the analysis. These trends are also confirmed on both the original samples, and on the ``purest'' one obtained after selecting only galaxies with SN([NII])$<$3.

\item At each redshift more massive galaxies present a median H$\beta$ index lower than less massive ones, confirming a mass-downsizing scenario for which more massive systems have assembled their stars earlier and faster. This result has been demonstrated not to depend on a selection effect due to the different velocity dispersions and metallicities of the samples; also in this case, this trend is found in both the original and the ``purest'' samples.

\item The comparison with SPS models highlights an inconsistency with observable data, for which observed galaxies appear, in most cases, to be older than the age of the Universe at the given redshifts. These differences are greater than $\Delta$H$\beta \sim 0.1$ \AA\, and seems not reconcilable with any possible further re-selection of the samples. Only a stellar metallicity systematically higher than the one found in these samples ($Z/Z_{\sun} \sim 2 $) may alleviate the tension between the data and models.

\item We tested the presence of a weak H$\beta$ emission line in our stacked spectra by using GANDALF code {and the emission corrections for the hydrogen features proposed by \cite{ServenWorthey2010}}. We find that the recovered emission lines are very uncertain, detected with a very low confidence level and are model dependent. The median emission equivalent width ranges from$\sim -0.1$ to $\sim-0.27$ \AA .

\item We also find that in the stacked spectrum obtained from the ``purest'' sample with SN([NII])$<$3, there is a residual [NII] emission line contamination, even if very weak (EW$\sim -0.26$).

\item Throughout simulations, we demonstrate that it exists a threshold limit below which an emission line component within the H$\beta$ absorption features would not be detectable; this threshold depends on the FWHM of the line, being EW(H$\beta$) $\leq$ -0.9 \AA\ or $\leq$ -0.2 \AA\, respectively for a broader and narrower FWHM (characteristic of star formation or AGN activity). We note that this EW is compatible with the observed offset between models and data.

\end{itemize}

In order to obtain a quantitative estimate of the age of formation of the galaxies and to test the feasibility of using this index as a ``cosmic chronometers" indicator, the observed H$\beta$-z relations should be calibrated on SPS models. However, the dependence of the normalization and of the slope of these relations on the different possible selections, and the apparent inconsistency with theoretical models, makes H$\beta$ an index that is difficult to rely on to estimate both absolute and relative ages, and this, as discussed, is due to a possible contamination of the line by an undetectable emission component. We discuss the possible candidates of ionization sources, finding a better agreement with post-AGB and slightly higher tension with AGN.

\

Despite this index has been historically identified as the best suited to constrain the age of a galaxy population, all the highlighted issues do not allow to use it for an accurate estimate of the absolute and differential age evolution of the stellar populations with redshift. Therefore, it is not possible to place constraints on the galaxies redshift of formation, and even the use of the H$\beta$-z relation in the ``cosmic chronometers'' approach appears inadequate.\

Finally, higher SN and resolution spectra may help in mitigating these problems during the selection phase, and to better disentangle a narrow emission line component for a less biased measurement. Another possible option is to study higher order Balmer lines (H$\gamma$, H$\delta$), as e.g. suggested by the work of \cite{Vazdekis1999},
since those lines should be less affected by an underlying emission component. This analysis will be further exploited in a following paper.

\section*{Acknowledgements}

         The authors would like to thank the referee, Guy Worthey, for useful suggestions. We also thank Marco Mignoli, Claudia Maraston, Daniel Thomas, Jonas Johansson, Marcella Brusa and Gianni Zamorani for the stimulating discussions.  We acknowledge Anna Gallazzi and Jarle Brinchmann for the SDSS data availability and their clarifications.  
AC is also grateful to Annalisa Citro and Salvatore Quai for sharing a preview of their results on SDSS stacked spectra analysis. Part of this work was supported by the INAF, Osservatorio Astronomico di Bologna and by the Dipartimento di Fisica e Astronomia, Universit\`a degli Studi di Bologna. 
 We acknowledge the grants ASI n. I/023/12/0  "Attivit\`a relative alla fase B2/C
per la missione Euclid" and PRIN MIUR 2010-2011 "The dark Universe and the cosmic evolution of baryons: from current surveys to Euclid". 
ACi, LP and MM acknowledge the PRIN MIUR 2015 "Cosmology and Fundamental Physics:
illuminating the Dark Universe with Euclid".








%
%


\bsp	
\label{lastpage}
\end{document}